\documentclass[12pt]{article}
\def\beq{\begin{equation}}
\def\eeq{\end{equation}}
\def\bea{\begin{eqnarray}}
\def\eea{\end{eqnarray}}
\def\bq{\begin{quote}}
\def\eq{\end{quote}}

\def\gappeq{\mathrel{\rlap {\raise.5ex\hbox{$>$}}
{\lower.5ex\hbox{$\sim$}}}}

\def\lappeq{\mathrel{\rlap{\raise.5ex\hbox{$<$}}
{\lower.5ex\hbox{$\sim$}}}}

\begin{document}
\topmargin -0.5cm
\oddsidemargin -0.3cm
\vspace*{5mm}
\begin{center}
{\bf On the uniqueness of solutions to gauge covariant Poisson equations with compact Lie algebras} \\
\vspace*{1.5cm} 
{\bf Christofer Cronstr\"{o}m}$^{*)}$ \\
\vspace{0.3cm}
Department of Physical Sciences, Theoretical Physics Division\\
FIN-00014 University of Helsinki, Finland \\
\vspace{0.5cm}
 
\vspace*{3cm}  
{\bf ABSTRACT} \\

\end{center}

It is shown, under rather general smoothness conditions on the gauge potential, which takes values in an 
arbitrary semi-simple compact Lie  algebra ${\bf g}$, that if a (${\bf  g}$-valued) solution to the gauge covariant Laplace 
equation exists, which vanishes at spatial infinity, in the cases of 1,2,3,... space dimensions, then the solution is 
identically zero. This result is also valid if the Lie algebra is merely compact. Consequently, a solution to the gauge covariant Poisson equation is uniquely determined by its asymptotic
radial limit at spatial infinity. In the cases of one or two space dimensions a related result is proved, namely that if a solution 
to the gauge covariant Laplace equation exists, which is unbounded at spatial infinity, but with a certain dimension-dependent 
condition on the asymptotic growth of its norm, then the solution in question is a covariant constant.

\vspace*{5mm}
\noindent

\vspace*{1cm} 

\vspace*{2cm} 
\noindent 
$^{*)}$ e-mail address: Christofer.Cronstrom@Helsinki.fi.  
\vspace*{1cm}
\vfill\eject


\setcounter{page}{1}
\pagestyle{plain}

\section{Introduction}
Let ${\bf  g}$ denote a semi-simple compact Lie algebra and let $\Phi$ and $\cal F$ denote two ${\bf  g}$-valued
functions on an $n$-dimensional Euclidean space $R^{n}$. Points in $R^{n}$ are denoted by 
$x := (x^{1}, x^{2},...,x^{n})$.  The covariant Poisson equation is then the following,
\beq
\sum_{k=1}^{n}\nabla_{k}(A)\nabla^{k}(A) \Phi(x) = {\cal F}(x),
\label{eq:Poiss}
\eeq
where $\nabla_{k}(A)$ denotes the following gauge-covariant gradient,
\beq
\nabla_{k}(A) := \partial_{k} + i[A_{k},~~] \equiv \frac{\partial}{\partial x^{k}} + i[A_{k}(x),~~].
\label{eq:covgr}
\eeq
In the definition (\ref{eq:covgr}) above, the symbol $A := (A_{1}(x), A_{2}(x),...,A_{n}(x))$  is a ${\bf g}$-valued
gauge potential \cite{Yang}, and the symbol $[~~,~~]$ stands for the commutator
of ${\bf g}$-valued quantities. The covariant gradient with an upper index $k$, $\nabla^{k}(A)$, is defined as follows,
\beq
\nabla^{k}(A):= - \nabla_{k}(A).
\label{eq:uppcov}
\eeq
Other quantities with upper space indices are defined similarly; {\em viz}. if $V_{k}$ is any quantity with a lower (subscript)
space index then its counterpart with an upper (superscript) space index $V^{k}$ is given as follows,
\beq
V^{k} := - V_{k}, \: k = 1, 2, ..., n.
\label{eq:contra}
\eeq

The covariant Poisson equation (\ref{eq:Poiss}) is a second order elliptic system of partial differential equations 
which together with appropriate boundary conditions is supposed to determine the quantity $\Phi(x)$, when the inhomogeneous 
term ${\cal F}(x)$ and gauge potential $A(x)$, respectively, are given. These quantities are supposed to satisfy appropriate 
regularity conditions. We will return to the question of regularity conditions subsequently.

The Lie algebra ${\bf g}$ is specified by the (real-valued) structure constants $f_{ab}^{~~c}$ in the commutation 
relations of the Lie algebra generators $T_{a}$, $a = 1, 2, ..., dim\, {\bf  g}$,
\begin{equation}
[T_{a}, T_{b}] = if_{ab}^{~~c}T_{c}, \; a, b = 1, 2, ...,  dim\, {\bf  g}. 
\label{eq:liealg}
\end{equation}    
Here  and in what follows, summation of repeated Lie algebra indices, i.e. letters from the beginning of the latin alphabet, over 
the range $1, 2, ..., dim\, {\bf  g}$ is implied, unless otherwise stated.

From now on we will identify the Lie algebra generators with some specific Hermitean matrix representatives. For the 
purposes in this paper it does not matter which representation is chosen; in physical applications the choice of 
representations is related to the question of what types of other fields are coupled to the gauge fields, a question 
which is not our concern here.  

Any ${\bf  g}$-valued quantity is in the linear span of the generators $T_{a}$, thus
\beq
\Phi(x) = \Phi^{a}(x)T_{a}, \: {\cal F}(x) =  {\cal F}^{a}(x)T_{a}, \; A_{k}(x) = A^{a}_{k}(x)T_{a},
\label{eq:linsp}
\eeq
where all the components $\Phi^{a}$, ${\cal F}^{a}$ and $A^{a}_{k}$ are {\em real}. 

The covariant Poisson equation (\ref{eq:Poiss}) can naturally be written as a system of partial differential equations
for the real-valued components $\Phi^{a}$. Using the commutator algebra (\ref{eq:liealg}) one readily obtains the following
system of equations,
\beq
\partial_{k}\partial^{k}\Phi^{a} - 2f_{bc}^{~~a}A^{kb}\partial_{k}\Phi^{c} - f_{bc}^{~~a}(\partial_{k}A^{kb})\Phi^{c}
+ f_{bd}^{~~a}f_{ec}^{~~d}A_{k}^{b}A^{ke} \Phi^{c} = {\cal F}^{a}.
\label{eq:Poiscomp}
\eeq
In the equation above summation over repeated space indices, one lower and one upper, is also implied.

For our purposes the compact matrix notation in Eq. (\ref{eq:Poiss}) is preferable, and this will be used in what follows.

We will state and prove our main uniqueness theorems in detail for the equation (\ref{eq:Poiss}) in Sections {\bf 3} and {\bf 4}
below. However before that we collect in the next section a number of facts related to gauge transformations and inner products 
for semi-simple compact Lie algebras which we need in the proofs.

\section{On gauge transformations and Lie algebra inner products}

Let $G$ denote the Lie group corresponding to the compact or semi-simple compact Lie algebra defined by the equations (\ref{eq:liealg}) 
and let $\omega$ denote a mapping from $R^{n}$ to $G$. In the vicinity of the unit element in $G$ any such $G$-valued element $\omega$ 
can be obtained by exponentiating the Lie algebra, i.e. 
\beq
\omega(x) = e^{\textstyle i\alpha^{a}T_{a}},
\label{eq:group}
\eeq
where the parameters $\alpha^{a}(x), a = 1, 2, ...,  dim\, {\bf  g}$ are sufficiently smooth real-valued functions which vary 
over a finite range.

A gauge transformation of a gauge potential $A \stackrel{\omega}{\rightarrow} A^{\omega}$ is then defined as follows for any 
sufficiently smooth $G$-valued function $\omega(x)$,
\begin{equation}
A_{k}(x)  \stackrel{\omega}{\longrightarrow} A^{\omega}_{k}(x) = \omega^{-1}(x) A_{k}(x) \omega(x)
+ i(\partial_{k}\omega^{-1}(x))\omega(x). 
\label{eq:gtrA}
\end{equation}
Let $U(x)$ be any differentiable ${\bf  g}$-valued function and $\omega(x)$ likewise any differentiable $G$-valued function.
Using Eq. (\ref{eq:gtrA}) one readily shows that
\beq
\omega^{-1}(x)\left (\nabla_{k}(A)U(x)Ê\right )\omega(x) = \nabla_{k}(A^{\omega})\left (\omega^{-1}(x)U(x)\omega(x)\right ).
\label{eq:simtr}
\eeq
Applying the transformation (\ref{eq:simtr}) to the gauge covariant Poisson equation (\ref{eq:Poiss}) one obtains
\beq
\sum_{k=1}^{n}\nabla_{k}(A^{\omega})\nabla^{k}(A^{\omega})\left (\omega^{-1}(x) \Phi(x)\omega(x)\right ) = \omega^{-1}(x){\cal F}(x)\omega(x),
\label{eq:Poitr}
\eeq
a result which will be used later.

If our Lie algebra is semi-simple and compact we use the following quantities $h_{ab}$ as our Lie algebra metric. The quantities $h_{ab}$
are defined in terms of the structure constants $f_{ab}^{~~c}$ as follows,
\beq
h_{ab} = -f_{ab'}^{~~c'}f_{bc'}^{~~b'}.
\label{eq:Killing}
\end{equation}
This is the so-called Killing form multiplied with $-1$ for convenience. It is known \cite{Hamer}, that the form (\ref{eq:Killing})
is non-degenerate, and furthermore positive definite, if and only if the Lie algebra is semi-simple and compact. The form 
(\ref{eq:Killing})  thus has an inverse, 
which we denote by $h^{ab}$,
\beq
h^{ab}h_{bc} = \delta^{a}_{~c}.
\label{eq:uppK}
\eeq
The form $h_{ab}$ ($h^{ab}$) is used to lower (raise) Lie algebra indices. 

For any two ${\bf g}$-valued quantities $U = U^{a}T_{a}$ and $V = V^{a}T_{a}$ we thus define their inner product $(U, V)$ as follows,
\beq
(U, V) = h_{ab}U^{a}V^{b}.
\label{eq:innp}
\eeq
The inner product (\ref{eq:innp}) is invariant under the adjoint action of the of the group, i.e.,
\beq
(U, V) = (\omega^{-1} U \omega, \omega^{-1} V \omega), \forall \omega \in G.
\label{eq:invp}
\eeq
The norm $||\;\;||_{g}$ of any Lie algebra valued quantity $U$ is defined in terms of the inner product,
\beq
||U||_{g} := \sqrt{(U, U)}.
\label{eq:gnorm}
\eeq

We also need the fact that for any three ${\bf g}$-valued quantities $U, V$ and $W$ the quantity $(U, [V, W])$ is cyclically 
symmetric,
\beq
(U, [V, W]) = (V, [W, U]) = (W, [U, V])
\label{eq:cycl}
\eeq
The equations (\ref{eq:cycl}) follow directly from the fact that the quantities $f_{abc}$ are antisymmetric under the interchange 
of any indices.

An immediate consequence of Eqs. (\ref{eq:cycl}) is the following useful identity,
\beq
\partial_{k}(U, V) = (\nabla_{k}(A)U, V) + (U, \nabla_{k}(A)V).
\label{eq:parA}
\eeq
This identity is valid for any two differentiable ${\bf g}$-valued quantities $U$ and $V$ and a smooth gauge potential $A$.

We finally note that one may drop the condition of semi-simplicity above and require only the condition of compactness of the Lie algebra.
One is still then guaranteed the existence of an inner product $(\;\;, \;\;)$ in the Lie algebra which is positive definite
and satisfies the condition (\ref{eq:cycl}) of cyclic symmetry \cite{Barut}. In the considerations below one only needs these 
properties of the Lie algebra inner product.  The results derived below are therefore also valid for the case of only compact Lie algebras,
and not only for the case of {\em both} semi-simple  and compact Lie algebras.

\section{Uniqueness theorems in the cases $n = 1$ and $n = 2$} 

\subsection{The one-dimensional case}

In the case of one space dimension, the gauge-covariant Poisson equation (\ref{eq:Poiss}) is fairly trivial. We 
nevertheless give a brief analysis also of this case for completeness. In one space dimension Eq. (\ref{eq:Poiss}) 
becomes the following,
\beq
\nabla_{1}(A)\nabla^{1}(A) \Phi(x) = {\cal F}(x),
\label{eq:Pois1}
\eeq
where we now use the notation $x$ in stead of $x^{1}$ for simplicity.

We then choose a $G$-valued quantity $\omega$ such that 
\beq
A_{1}^{\omega}(x) \equiv \omega^{-1}(x)A_{1}(x)\omega(x) + i\left (\frac{d}{dx} \omega^{-1}(x)\right )\omega(x) = 0.
\label{eq:ahom}
\eeq
The condition (\ref{eq:ahom}) is equivalent to the differential equation
\beq
\frac{d}{dx} \omega(x) = -i A_{1}(x)\omega(x).
\label{eq:diffo}
\eeq
For any smooth gauge potential $A_{1}(x)$, the (matrix) differential equation (\ref{eq:diffo}) has a $G$-valued solution, 
determined apart from a constant initial value at some appropriate point in $R^{1}$. Then, applying a transform of the type 
given in Eq. (\ref{eq:Poitr}) to Eq. (\ref{eq:Pois1}), with the quantity $\omega$ determined by Eq. (\ref{eq:diffo}), one obtains,
\beq
\frac{d^{2}}{dx^{2}} \left (\omega^{-1}\Phi \omegaÊ\right ) =  \left (\omega^{-1}{\cal F} \omegaÊ\right ).
\label{eq:Poi12nd}
\eeq

The question of uniqueness of the solution(s) to Eq. (\ref{eq:Poi12nd}) is related to the number of solutions of the homogeneous
equation corresponding to Eq. (\ref{eq:Poi12nd}) under appropriate boundary conditions. The homogeneous equation in question,
\beq
\frac{d^{2}}{dx^{2}} \left (\omega^{-1}\Phi_{h} \omegaÊ\right ) = 0,
\label{eq:Poihom}
\eeq
has the general solution
\beq
\Phi_{h}(x)  =  \omega(x)\,\alpha\,Ê\omega^{-1}(x)\,x + \omega(x)\,\beta\,Ê\omega^{-1}(x),
\label{eq:solhom}
\eeq
where $\alpha$ and $\beta$ are arbitrary constant ${\bf g}$-valued quantities.

If it is required to find solutions $\Phi(x)$ to Eq. (\ref{eq:Pois1}), which is equivalent to Eq.(\ref{eq:Poi12nd}), which grow less rapidly 
than linearly with $x$ for large $x$, then the following boundary condition at infinity must hold,
\beq
\lim_{|x| \rightarrow \infty}  \frac{1}{x} ||\Phi(x)||_{g} = 0. 
\label{eq:asy1}
\eeq
Hence the difference of any two such solutions, which satisfies the homogeneous equation (\ref{eq:Poihom}), and which is of the form $\Phi_{h}$
given in Eq. (\ref{eq:solhom}) above, must satisfy the condition
\beq
\lim_{|x| \rightarrow \infty}  \frac{1}{x} ||\Phi_{h}(x)||_{g} = \lim_{|x| \rightarrow \infty}  \frac{1}{x} \sqrt{(\alpha,\alpha)x^{2} + 2(\alpha,\beta)x +(\beta,\beta)} = 0.
\label{eq:asyh}
\eeq
Thus one must set $\alpha = 0$ in Eq. (\ref{eq:solhom}). Hence, if a solution to Eq. (\ref{eq:Pois1}) exists, which satisfies the 
condition (\ref{eq:asy1}), then the solution is unique apart from an additive {\em covariant constant}, i.e. a solution to the 
homogeneous equation of the form
\beq
\Phi_{h}(x)  =   \omega(x)\,\beta\,Ê\omega^{-1}(x).
\label{eq:solhom2}
\eeq
The solution (\ref{eq:solhom2}) satisfies the condition of covariant constancy,
\beq
\nabla_{1}(A) \Phi_{h}(x) = 0
\eeq
as a consequence of the condition (\ref{eq:diffo}) which determines the quantity $\omega$.
 
Likewise, if one requires solutions which vanish at infinity, i.e. which satisfy the
conditions
\beq
\lim_{|x| \rightarrow \infty} ||\Phi(x)||_{g} = 0,
\label{eq:asy12}
\eeq
then one must have $\alpha = \beta = 0$ in Eq. (\ref{eq:solhom}). Thus, if a solution to Eq. (\ref{eq:Pois1}) exists, which satisfies
the condition (\ref{eq:asy12}), then the solution is unique. 

We can summarise the discussion above as follows:

\noindent
{\em Theorem 3.1}

\noindent
Assume that the gauge-covariant Poisson equation in one space dimension, Eq. (\ref{eq:Pois1}), has a solution in the classic sense,
i.e. a solution which is twice 
continuously differentiable in any finite interval in $R^{1}$, for a given smooth gauge potential $A_{1}(x)$ and for a given 
inhomogeneous term ${\cal F}(x)$, satisfying appropriate smoothness and asymptotic conditions in $R^{1}$. If such a solution grows 
less rapidly than linearly with increasing $x$, i.e. satisfies the condition (\ref{eq:asy1}), then the solution is unique apart from 
an additive covariant constant. Furthermore, if such a solution vanishes asymptotically in the sense of condition (\ref{eq:asy12}) 
then the solution is unique.

\subsection{The two-dimensional case}

We will now consider the gauge covariant Poisson equation (\ref{eq:Poiss}) in two-dimensional space $R^{2}$. We assume that this 
equation has solutions which are twice continuously differentiable in any finite domain in $R^{2}$, for a given smooth gauge
potential $(A_{1}(x^{1}, x^{2}), A_{2}(x^{1}, x^{2}))$ and a given inhomogeneous term ${\cal F}(x^{1}, x^{2})$, which is supposed to satisfy
appropriate smoothness and asymptotic conditions.  Let there be two such solutions $\Phi_{A}$ and $\Phi_{B}$, say. Then their
difference
\beq
\Phi_{C} := \Phi_{A} - \Phi_{B}
\label{eq:differ}
\eeq
satisfies the gauge covariant Laplace equation, i.e. the homogeneous equation,
\beq
\nabla_{k}(A)\nabla^{k}(A) \Phi_{C}(x) = 0,
\label{eq:Lap2}
\eeq
where summation over the space-index $k$ in the range $(1,2)$ is implied here and in what follows. We then consider the quantity
\beq
- \nabla^{2}(\Phi_{C}, \Phi_{C}) \equiv \partial_{k}\partial^{k}(\Phi_{C}, \Phi_{C}).
\label{eq:nabl2}
\eeq
Using the previously established formula (\ref{eq:parA}) and the equation (\ref{eq:Lap2}) above, one obtains
\beq
- \nabla^{2}(\Phi_{C}, \Phi_{C}) = 2(\nabla_{k}(A)\Phi_{C}, \nabla^{k}(A)\Phi_{C}).
\label{eq:nabeq}
\eeq
We then express the $\nabla^{2}$-operator in terms of polar coordinates $(r, \theta)$ in the plane $R^{2}$,
\beq
\nabla^{2} = \frac{\partial^{2}}{\partial r^{2}} + \frac{1}{r}\frac{\partial }{\partial r} + \frac{1}{r^{2}}\frac{\partial^{2}}{\partial \theta^{2}}.
\label{eq:pollap}
\eeq
Integrating the equation (\ref{eq:nabeq}) over the variable $\theta$ in the range $(0, 2\pi)$ one obtains,
\beq
\left (\frac{\partial^{2}}{\partial r^{2}} + \frac{1}{r}\frac{\partial }{\partial r}\right )\int_{0}^{2\pi} d\theta \,(\Phi_{C}, \Phi_{C}) = - 2\int_{0}^{2\pi} d\theta \, (\nabla_{k}(A)\Phi_{C}, \nabla^{k}(A)\Phi_{C}).
\label{eq:cdiffeq}
\eeq
We now consider the equation (\ref{eq:cdiffeq}) above as an ordinary differential equation for the quantity $\int d\theta\,(\Phi_{C}, \Phi_{C})$
as if the right hand side of the equation were a known quantity. The general solution to this equation is then the following,
\bea
\label{eq:Csol2}
\int_{0}^{2\pi} d\theta \,(\Phi_{C}, \Phi_{C})  & =  &\alpha\log\,r + \beta   \nonumber\\
& - & 2\, (\log\, r) \int_{0}^{r} dr'\,r'  \int_{0}^{2\pi} d\theta \, (\nabla_{k}(A)\Phi_{C}, \nabla^{k}(A)\Phi_{C}) \\
& + & 2 \int_{0}^{r} dr' \, r' (\log\,r') \int_{0}^{2\pi} d\theta \, (\nabla_{k}(A)\Phi_{C}, \nabla^{k}(A)\Phi_{C}), \nonumber
\eea
where $\alpha$ and $\beta$ are real constants.

In view of the assumed smoothness of the gauge potential $A$ and assumed regularity of the solution $\Phi_{C}$ to the original 
equation (\ref{eq:Lap2}), in particular in the vicinity of the origin $r = 0$ in $R^{2}$, one readily concludes that the terms
involving integrals in the expression (\ref{eq:Csol2}) can not generate logarithmic singularities at $r = 0$ which could cancel the 
term $\alpha\log\,r$ in that expression. The logarithmic singularity in (\ref{eq:Csol2})  must be expelled, and this can then only be 
achieved by setting $\alpha = 0$. Hence, the general solution to the equation (\ref{eq:cdiffeq}) when the regularity conditions
at $r = 0$ are taken into account, is
\beq
\int_{0}^{2\pi} d\theta \,(\Phi_{C}, \Phi_{C}) =  \beta - 2\, (\log\, r) \int_{0}^{r} dr'\,r' \left (1 - \frac{\log\,r'}{\log\,r} \right ) \int_{0}^{2\pi} d\theta \, (\nabla_{k}(A)\Phi_{C}, \nabla^{k}(A)\Phi_{C}),
\label{eq:finCsol2}
\eeq
where we now can identify the constant $\beta$ as follows,
\beq
\beta = \left [\int_{0}^{2\pi} d\theta \,(\Phi_{C}, \Phi_{C}) \right ]_{r=0} = 2\pi ||\Phi_{C}(0)||_{g}^{2}.
\label{eq:beta2}
\eeq

We then consider the asymptotic properties of the solutions $\Phi_{C}$. Let us impose the condition
\beq
\lim_{r \rightarrow \infty} \frac{1}{\log \,r} \,\int_{0}^{2\pi} d\theta \,(\Phi_{C}, \Phi_{C}) = 0.
\label{eq:asympR21}
\eeq
Then Eq. (\ref{eq:finCsol2}) implies that
\beq
\lim_{r\rightarrow \infty}\, \int_{0}^{r} dr'\,r' \left (1 - \frac{\log\,r'}{\log\,r} \right ) \int_{0}^{2\pi} d\theta \, (\nabla_{k}(A)\Phi_{C}, \nabla^{k}(A)\Phi_{C}) = 0.
\label{eq:condasyR21}
\eeq
The vanishing of the integral (\ref{eq:condasyR21}) in the limit $r \rightarrow \infty$ implies that the integrand must vanish, since 
the integrand is non-positive (for $r > 1$) in view of convention (\ref{eq:uppcov}) and the positive definiteness of the Lie algebra 
inner product. But then, necessarily,
\beq
\nabla^{k}(A)\Phi_{C} = 0,\;k = 1, 2,
\label{eq:covcon2}
\eeq
i.e. $\Phi_{C}$ is a covariant constant.  

We can  thus infer from the  result above, that if the gauge covariant Poisson equation in $R^{2}$
has a twice continuously differentiable solution $\Phi$, such that $(\Phi, \Phi)$ is dominated by $\log \, r$ for large values of $r$, 
or more precisely, if the solution satisfies an  asymptotic condition of the form (\ref{eq:asympR21}), then the solution is unique 
apart from an additive covariant constant. Let us note however, that the  existence of a non-zero covariant constant in two (or more) 
dimensions places certain restrictions on the gauge potential $A$, which are related to the internal holonomy group \cite{Loos}.

If one requires a more stringent asymptotic condition on the solutions to the gauge covariant Laplace equation than condition
(\ref{eq:asympR21}), namely the condition
\beq
\lim_{r \rightarrow \infty} \int_{0}^{2\pi} d\theta \,(\Phi_{C}, \Phi_{C}) \equiv \lim_{r \rightarrow \infty} \int_{0}^{2\pi} d\theta \,||\Phi_{C}||^{2}_{g} = 0,
\label{eq:asympR22}
\eeq
then Eqns. (\ref{eq:finCsol2}) and (\ref{eq:beta2}) imply as before that Eq. (\ref{eq:covcon2}) is in force, but also that
\beq
\left [\int_{0}^{2\pi} d\theta \,(\Phi_{C}, \Phi_{C}) \right ]_{r=0} = 0.
\label{eq:Phiorig}
\eeq
Taken together, Eqns. (\ref{eq:covcon2}) and (\ref{eq:Phiorig}) imply that
\beq
\Phi_{C}(x^{1}, x^{2}) \equiv 0.
\label{eq:zero2}
\eeq

We have now demonstrated that the only solution to the gauge covariant Laplace equation in two dimensions which vanishes at
spatial infinity in the sense given by Eq. (\ref{eq:asympR22}), is the identically vanishing solution. This gives rise to a uniqueness
theorem for such solutions $\Phi$ to the inhomogeneous gauge covariant Poisson equation, which have a given asymptotic behaviour $\Phi^{as}$
at spatial infinity, or more precisely, which satisfy a boundary condition of the form
\beq
\lim_{r \rightarrow \infty} \int_{0}^{2\pi} d\theta \,||\Phi - \Phi^{as}||_{g} = 0.
\label{eq:asynoh}
\eeq
Namely, suppose that there exist two solutions, $\Phi_{A}$ and $\Phi_{B}$, say, to the gauge covariant Poisson equation in two dimensions,
which satisfy a boundary condition of the form (\ref{eq:asynoh}) with some appropriate given asymptotic function $\Phi^{as}$. Then their 
difference $\Phi_{C}$ satisfies the gauge covariant Laplace equation (\ref{eq:Lap2}) and the following boundary condition,
\bea
\label{eq:difasy}
\lefteqn{0 \leq \lim_{r\rightarrow \infty} \int_{0}^{2\pi} d\theta \, ||\Phi_{C}||^{2}_{g} = \lim_{r\rightarrow \infty} \int_{0}^{2\pi} d\theta \, ||\Phi_{A} -  \Phi_{B}||^{2}_{g}} \nonumber\\
& = & \lim_{r\rightarrow \infty} \int_{0}^{2\pi} d\theta \, ||(\Phi_{A} - \Phi^{as}) - (\Phi_{B} - \Phi^{as})||^{2}_{g} \\
& \leq & \lim_{r\rightarrow \infty} \int_{0}^{2\pi} d\theta \,(||\Phi_{A} - \Phi^{as}||_{g} + ||\Phi_{B} - \Phi^{as}||_{g})^{2} = 0.\nonumber
\eea
But then, in accordance with the reasoning above, Eq. (\ref{eq:zero2}) must be in force, i.e. $\Phi_{A} \equiv \Phi_{B}$.
Thus, if a solution to the gauge covariant Poisson equation in $R^{2}$ exists which has a given asymptotic limit in the sense given
in Eq. (\ref{eq:asynoh}), then the solution is unique.

We summarise the results obtained above on the gauge covariant Poisson equation on $R^{2}$ as {\em Theorem 3.2.} below.

\noindent
{\em Theorem 3.2}

\noindent
Assume that the gauge-covariant Poisson equation in two space dimensions has a solution in the classic sense,
i.e. a solution which is twice continuously differentiable in any finite domain in $R^{2}$, for a given smooth gauge potential 
$(A_{1}(x^{1}, x^{2}), A_{2}(x^{1}, x^{2}))$ and for a given inhomogeneous term ${\cal F}(x^{1}, x^{2})$, satisfying appropriate 
smoothness and asymptotic conditions in $R^{2}$. 

If the squared norm $||\Phi||^{2}_{g}$ of such a solution grows less rapidly than logarithmically with increasing distance from the origin 
in $R^{2}$, i.e. satisfies the condition (\ref{eq:asympR21}), then the solution is unique apart from an additive covariant constant. 

Furthermore, if such a solution has a given asymptotic radial limit $\Phi^{as}$ in the sense of condition (\ref{eq:asynoh}),
then the solution is unique.

\section{Uniqueness theorem in the cases $n \geq 3$}

The analysis of the uniqueness of solutions to the gauge covariant Poisson equation (\ref{eq:Poiss}) in the cases $n \geq 3$
proceeds much in the same way as in the two-dimensional case considered in the previous subsection.  Thus we assume that the 
equation (\ref{eq:Poiss}) has at least two solutions $\Phi_{A}(x)$ and $\Phi_{B}(x)$, say, which are twice continuously differentiable
in any finite domain in $R^{n}$, for any given sufficiently smooth gauge potential $(A_{1}(x), A_{2}(x), \ldots, A_{n}(x))$ and 
given inhomogeneous term ${\cal F}(x)$, which satisfies appropriate smoothness and asymptotic conditions in $R^{n}$. We consider
the difference $\Phi_{C}(x)$ of these solutions,
\beq
\Phi_{C}(x) = \Phi_{A}(x) - \Phi_{B}(x).
\label{eq:differ3}
\eeq
The function $\Phi_{C}(x)$ satisfies the  gauge covariant Laplace equation,
\beq
\sum_{k=1}^{n} \nabla_{k}(A)\nabla^{k}(A) \Phi_{C} = 0
\label{eq:Lap3}
\eeq
as well as such regularity and asymptotic conditions which follow from those conditions of a similar nature which are supposed to be valid for 
the solutions $\Phi_{A}(x)$ and $\Phi_{B}(x)$. In the equation (\ref{eq:Lap3}) we have for clarity reinstated the explicit summation over the space index $k$ and continue to use this 
notation below.

Again, using the previously established formula (\ref{eq:parA}) and the equation (\ref{eq:Lap3}) above, one obtains
\beq
\nabla^{2}(\Phi_{C}, \Phi_{C}) =  - 2\,\sum_{k=1}^{n}(\nabla_{k}(A)\Phi_{C}, \nabla^{k}(A)\Phi_{C}).
\label{eq:nabeq3}
\eeq
We then use spherical coordinates \cite{Vilen} on $R^{n}$,
\begin{eqnarray}
\label{eq:sphcor}
x^{1} & = & r\sin\theta_{n-1}\ldots\sin\theta_{2}\sin\theta_{1}\nonumber\\
x^{2} & = & r\sin\theta_{n-1}\ldots\sin\theta_{2}\cos\theta_{1}\nonumber\\
& \vdots &\\
x^{n-1} & = & r\sin\theta_{n-1}\cos\theta_{n-2}\nonumber\\
x^{n} & = & r\cos\theta_{n-1}\nonumber
\end{eqnarray}
where
\begin{equation}
r \geq 0,\;0 \leq \theta_{1} < 2\pi, \; 0 \leq \theta_{k} \leq \pi,\; k \neq 1.
\label{eq:doma}
\end{equation}
The Laplace operator $\nabla^{2}$ expressed in terms of the spherical coordinates above is as follows,
\begin{eqnarray}
\label{eq:sphlap}
\nabla^{2}& = &\frac{1}{r^{n-1}}\frac{\partial}{\partial r}r^{n-1}\frac{\partial}{\partial r}\\
          & + &\frac{1}{r^{2}\sin^{n-2}\theta_{n-1}}\frac{\partial}{\partial \theta_{n-1}}\sin^{n-2}\theta_{n-1}\frac{\partial}{\partial \theta_{n-1}}\nonumber\\
          & + & \frac{1}{r^{2}\sin^{2}\theta_{n-1}\sin^{n-3}\theta_{n-2}}\frac{\partial}{\partial \theta_{n-2}}\sin^{n-3}\theta_{n-2} \frac{\partial}{\partial \theta_{n-2}}\nonumber\\
          & + & \ldots + \frac{1}{r^{2}\sin^{2}\theta_{n-1} \ldots \sin^{2}\theta_{2}}\frac{\partial^{2}}{\partial \theta_{1}^{2}}.\nonumber
\end{eqnarray}
We still need the invariant (normalised) measure $d\Omega_{n}$ on the sphere $S^{n-1}$ in terms of the spherical coordinates above,
\begin{equation}
d\Omega_{n} = \frac{\Gamma\left (\frac{n}{2}\right )}{2\pi^{\frac{n}{2}}}\sin^{n-2}\theta_{n-1}\ldots \sin\theta_{2}d\theta_{1}\ldots d\theta_{n-1}.
\label{eq:meas}
\end{equation} 
Using the formulae (\ref{eq:sphlap}) and (\ref{eq:meas}) above, one obtains the following result from Eq. (\ref{eq:nabeq3}),
\beq 
\frac{1}{r^{n-1}}\frac{\partial}{\partial r}\left (r^{n-1}\frac{\partial }{\partial r}\,\int d\Omega_{n}\,(\Phi_{C}, \Phi_{C})\right ) = -2\,\sum_{k=1}^{n} \int d\Omega_{n}\,(\nabla_{k}(A)\Phi_{C}, \nabla^{k}(A)\Phi_{C}),
\label{eq:diffeq3}
\eeq
where the integration over the angular variables $\theta_{k}, k = 1,2,\ldots,n-1$ is over the complete range specified in (\ref{eq:doma})
above.    
The equation (\ref{eq:diffeq3}) is now considered as an ordinary differential equation for the quantity $\int d\Omega_{n}\, (\Phi_{C}, \Phi_{C})$,
under the assumption that the right hand side of Eq.  (\ref{eq:diffeq3})  is a known quantity. The general solution to Eq.  (\ref{eq:diffeq3}) is then the
following,
\bea
\label{eq:solu3}
\int d\Omega_{n}\,(\Phi_{C}, \Phi_{C}) & = & \alpha r^{2-n} + \beta + \nonumber\\
& + &\frac{2\,r^{2-n}}{n-2} \int_{0}^{r}dr'\,r'^{n-1} \int d\Omega_{n}\,\sum_{k=1}^{n}(\nabla_{k}(A)\Phi_{C}, \nabla^{k}(A)\Phi_{C})\\
& - & \frac{2}{n-2} \int_{0}^{r}dr'\,r' \int d\Omega_{n}\,\sum_{k=1}^{n}(\nabla_{k}(A)\Phi_{C}, \nabla^{k}(A)\Phi_{C}), \nonumber
\eea
where $\alpha$ and $\beta$ are so far undetermined real constants. 

We now recall that the solutions $\Phi_{C}$ are supposed to be twice differentiable in any finite domain in $R^{n}$, in particular in the 
vicinity of the origin $r = 0$ in $R^{n}$. Moreover, the gauge potential $A$ is supposed to be smooth, in particular near $r = 0$ in 
$R^{n}$. From these conditions follow the  estimates below, valid near $r = 0$, 
\beq
\left | \int_{0}^{r} dr'\,r'^{n-1} \int d\Omega_{n}\,\sum_{k=1}^{n}(\nabla_{k}(A)\Phi_{C}, \nabla^{k}(A)\Phi_{C}) \right | = O(r^{n})
\label{estn3}
\eeq
and
\beq
\left | \int_{0}^{r} dr'\,r' \int d\Omega_{n}\,\sum_{k=1}^{n}(\nabla_{k}(A)\Phi_{C}, \nabla^{k}(A)\Phi_{C}) \right | = O(r^{2}).
\label{est23}
\eeq
From the esimates (\ref{estn3}), (\ref{est23}) and the equation (\ref{eq:solu3}) it then follows that
\beq
\int d\Omega_{n}\,(\Phi_{C}, \Phi_{C})  = \alpha r^{2-n} + \beta  + O(r^{2})
\label{totest}
\eeq
in the vicinity of $r = 0$. But the solution $\Phi_{C}$ is supposed to be regular, in particular near $r = 0$. The singularity
at $r = 0$ which appears to be present in the general solution (\ref{eq:solu3}) must be made to disappear, and this can only
happen if
\beq
\alpha = 0
\label{eq:alph0}
\eeq
in Eq. (\ref{eq:solu3}), in accordance with the estimate (\ref{totest}). Then it also follows that
\beq
\beta = \left [\int d\Omega_{n}\,(\Phi_{C}, \Phi_{C}) \right ]_{r=0} = ||\Phi_{C}(0)||_{g}^{2}.
\label{betan3}
\eeq
Using the conditions (\ref{eq:alph0}) and (\ref{betan3}), one then finally obtains the following result,
\bea
\label{CCfin3}
\lefteqn{\int d\Omega_{n}\,(\Phi_{C}, \Phi_{C})  =  ||\Phi_{C}(0)||_{g}^{2}} \\
& - &\frac{2}{n-2}\int_{0}^{r} dr'\,r'\left (1 - \left (\frac{r'}{r} \right )^{n-2} \right ) \int d\Omega_{n}\,\sum_{k=1}^{n}(\nabla_{k}(A)\Phi_{C}, \nabla^{k}(A)\Phi_{C}). \nonumber
\eea
Let us emphasize that the equation (\ref{CCfin3}) is a relation which is valid for any twice differentiable solution $\Phi_{C}$ to the gauge 
covariant Laplace equation (\ref{eq:Lap3}) with a smooth gauge potential $A$.

Assume now the following boundary condition at spatial infinity for the solution $\Phi_{C}$ to the gauge covariant  Laplace 
equation (\ref{eq:Lap3}), 
\beq
\lim_{r\rightarrow \infty} \int d\Omega_{n} (\Phi_{C}, \Phi_{C}) = 0.
\label{Casymp3}
\eeq
In view of the convention (\ref{eq:uppcov}) and the positive definiteness of the inner product $(\;,\;)$, the condition (\ref{Casymp3}) 
and the relation (\ref{CCfin3}) together imply that
\beq
\nabla^{k}(A)\Phi_{C} = 0, \; k = 1,2,\dots, n,
\label{covco3}
\eeq
and that
\beq
||\Phi_{C}(0)||_{g}^{2} = 0.
\label{norm3=0}
\eeq
But the conditions (\ref{covco3}) and (\ref{norm3=0}) then finally imply that
\beq
\Phi_{C} \equiv 0.
\label{Consta3}
\eeq

We have thus shown that the only twice differentiable solution $\Phi_{C}$ to the gauge covariant Laplace equation (\ref{eq:Lap3}) with a 
smooth gauge potential $A$, which vanishes at spatial infinity in the sense of  the condition (\ref{Casymp3}), is the identically vanishing 
solution (\ref{Consta3}). 

The result above gives rise to a uniqueness theorem for the solutions to the gauge covariant  Poisson equation (\ref{eq:Poiss}) in a space
$R^{n}$ of $n \geq 3$ dimensions, just as in the two-dimensional case. Namely, consider the following boundary conditions,
\beq
\lim_{r \rightarrow \infty} \int d\Omega_{n} \,||\Phi - \Phi^{as}||_{g} = 0.
\label{eq:asyndim}
\eeq 
As has essentially already been demonstrated in the two-dimensional case, then the difference $\Phi_{C}$, Eq. (\ref{eq:differ3}), of any two 
solutions $\Phi_{A}$ and $\Phi_{B}$ satisfying the asymptotic condition (\ref{eq:asyndim}), vanishes at spatial infinity in the sense of Eq.
(\ref{Casymp3}). The difference in question also satisfies the gauge covariant Laplace equation (\ref{eq:Lap3}), and vanishes therefore 
identically, as has just been demonstrated above. Hence the solution to the gauge covariant Poisson equation (\ref{eq:Poiss}) is unique if one 
imposes the boundary conditions (\ref{eq:asyndim}).

We summarise the result above as the following theorem of uniqueness:

\noindent
{\em Theorem 4}.

\noindent
Assume that the gauge-covariant Poisson equation (\ref{eq:Poiss}) in $n \geq 3$ space dimensions has a solution in the classic sense,
i.e. a solution which is twice continuously differentiable in any finite domain in $R^{n}$, for a given smooth gauge potential 
$(A_{1}(x), A_{2}(x), \ldots, A_{n}(x))$ and for a given inhomogeneous term ${\cal F}(x)$, satisfying appropriate 
smoothness and asymptotic conditions in $R^{n}$. If, furthermore, such a solution has a given asymptotic limit $\Phi^{as}$  in the sense of 
condition (\ref{eq:asyndim}), then the solution is {\em unique}.

\section{Summary and discussion}

In this paper a  uniqueness theorem has been proved for the gauge covariant Poisson equation in n-dimensional space $R^{n}$.
The theorem has been obtained by considering the homogeneous counterpart of the Poisson equation in question, i.e. the gauge 
covariant Laplace equation. It has been shown that the only solution to the gauge covariant Laplace equation, which is twice continuously
differentiable in any finite domain in $R^{n}$, and which vanishes at infinity is the zero solution. This then proves the uniqueness
of that solution of the corresponding Poisson equation, which satisfies appropriate conditions of regularity and has a given asymptotic radial
limit.

In one- or two dimensions the asymptotic conditions  can be relaxed; it has been shown that in these cases the solutions, which may 
be unbounded at infinity, are unique apart from an additive covariant constant, even if one does not specify the asymptotic behaviour
of the solutions, but merely imposes certain specific limitations on the asymptotic growth of the solutions.  

We have not touched upon the question of existence of solutions in this paper. Such questions have been analysed in depth in the 
three-dimensional case in a recent paper by Salmela \cite{Salm}, who uses modern functional analytic methods in his study. This 
paper also contains references to physical applications of the gauge covariant Poissson equations in $R^{3}$. \\

\bigskip

{\bf Acknowledgements}

I wish to express my gratitude to Antti Salmela for useful and enjoyable discussions on the topics presented in this paper.
I am also indebted to Claus Montonen for important comments on the question of appropriate boundary conditions for gauge
covariant Poisson equations.

\end{document}